\documentclass[a4paper]{jpconf}
\usepackage{graphicx}
\begin{document}
\title{Liquid Xenon Detectors for Positron Emission Tomography}

\author{A. Miceli$^1$, P. Amaudruz$^1$,  F. Benard$^2$,  D. A. Bryman$^3$, L. Kurchaninov$^1$, J. P. Martin$^4$, A. Muennich$^1$,  F. Retiere$^1$, T. J. Ruth$^1$, V. Sossi$^3$, A. J. Stoessl$^3$}

\address{$^1$ TRIUMF, 4004 Wesbrook Mall, 
Vancouver, BC,  V6T~2A3, Canada}
\address{$^2$ BC Cancer Agency, 600 West 10th Avenue, Vancouver, BC, V5Z~4E6, Canada}
\address{$^3$ Department of Physics and Astronomy, 
University of British Columbia, 6224 Agricultural Road, Vancouver, BC, V6T~1Z1, Canada}
\address{$^4$ University of Montreal, CP 6128 Succursale Centre-Ville, Montreal, Quebec, H3C~3J7, Canada}

\ead{miceli@triumf.ca}

\begin{abstract}
PET is a functional imaging technique based on detection of annihilation photons following beta decay producing positrons. In this paper, we present the concept of a new PET system for preclinical applications consisting of a ring of twelve time projection chambers filled with liquid xenon viewed by avalanche photodiodes.  Simultaneous measurement of ionization charge and scintillation light leads to a significant improvement to spatial resolution, image quality, and sensitivity. Simulated performance shows that an energy resolution of $<10 \%$ (FWHM) and a sensitivity of 15$\%$ are achievable. First tests with a prototype TPC indicate position resolution  $<1$ mm (FWHM).
\end{abstract}

\section{Introduction}
Positron emission tomography (PET) is a functional imaging technique based on detection of gamma rays from positron annihilation. Liquid xenon (LXe) gamma ray detectors are excellent candidates for PET due to copious scintillation and ionization signals which are anti-correlated \cite{Kubota1978}.  Measuring both charge and light  leads to an improvement of energy resolution. Combined energy resolutions smaller than 4$\%$ FWHM at 662 keV have been reported experimentally \cite{Aprile07}. Gamma rays interacting with LXe produce large detectable signals due to the high ionization yield (64e-/keV at high E field) and light output (68 photons/keV at zero E field).   The fast scintillation light decay time (2.2ns) allows sub-ns timing resolution. Ionization charge can drift long distances in LXe with little spreading (20$\mu$m diffusion for 1 $\mu$s drift) \cite{Doke91} allowing sub-mm position resolution.  Good energy resolution, sub-ns timing resolution, and 3D sub-mm position resolution result in improved image contrast and spatial resolution, and reduced image artefacts. If a time projection chamber (TPC) is used, the position and energy of the each interaction can be measured and,  in case of multiple interactions, the interaction sequence can be reconstructed using Compton event reconstruction algorithms \cite{Oberlack00}. For high rates, the light signal can be used to roughly localize the position of interaction while the charge is drifting to the anode.  LXeTPC detectors have been successfully employed for particle physics, astrophysics, and dark matter detection  \cite{Angle08} \cite{Neilson09} \cite{Aprile01}. Their application to PET, however, remain to be demostrated.  
In this work we present a novel concept of a PET system for preclinical applications based on LXeTPC detectors. Simulated performace and first results  obtained with a sector of the LXe PET system will be discussed.

\section{LXe Micro-PET System}
\subsection{Concept} 
\begin{figure}[t]
\begin{minipage}[b]{18pc}
\includegraphics[width=18pc]{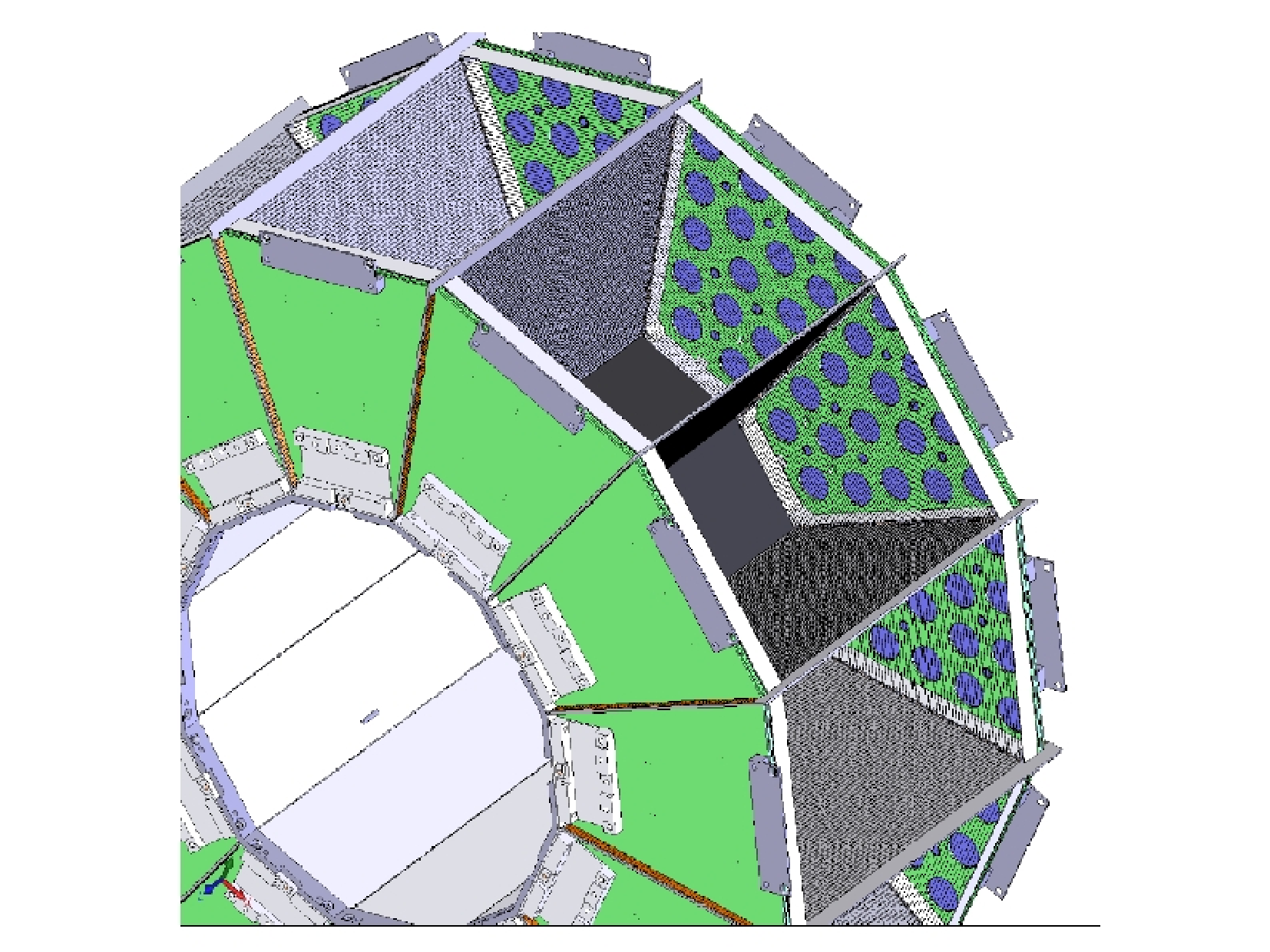}
\caption{\label{fig:PET_ring}LXe micro-PET ring.}
\end{minipage}  \hspace{1.5pc}
\begin{minipage}[b]{18pc}
\includegraphics[width=18pc]{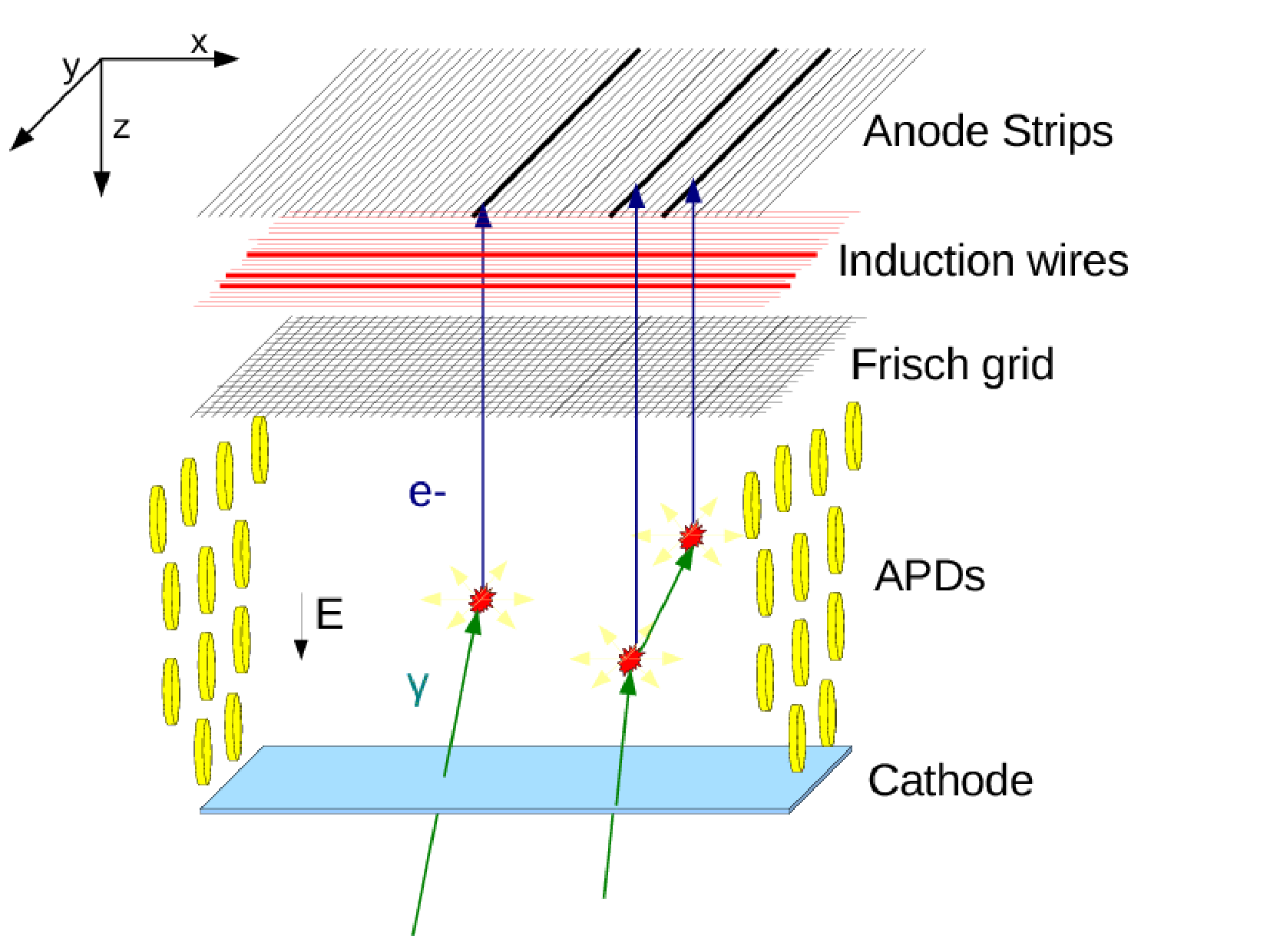}
\caption{\label{fig:principleOfOperation.ps}Sketch of the LXe PET detector.}
\end{minipage} 
\end{figure}

The proposed PET system for preclinical applications consists of twelve trapezoidal sectors arranged in a ring geometry placed in a vessel filled with LXe located inside a vacuum cryostat. Each sector consists of a TPC immersed in LXe viewed by two arrays of avalanche photodiodes (APDs) (Figure \ref{fig:PET_ring}).  The anode module of the TPC consists of a Frisch grid, a layer of induction wires, and an anode segmented into strips perpendicular to the wires. Annihilation photons entering the detector from the cathode side interact with the LXe producing ionization charge and scintillation light at 178 nm. The induction wires provide the y position of the interaction, whereas the anode strip which collects the charge provides the x coordinate. The  z-coordinate is calculated from the drift time and the drift velocity. The amplitude of the pulse on the anode strip is proportional to the energy deposited in the interaction. The scintillation light measured by the APDs provides the trigger both for readout of charge signals and coincidence measurements and it is used for fast localization of the position of the interaction at high rates. Figure \ref{fig:principleOfOperation.eps} shows a sketch of the LXe PET detector. Both scintillation light and ionization charge are used for energy resolution.  A Compton reconstruction algorithm is used to reconstruct the interaction sequence of multi-interaction events. The same algorithm can be used for the rejection of random coincidences and scatter. 

\subsection{Simulated Performance} 
The simulation of the LXe PET prototype was carried out with the Monte Carlo code Geant4 \cite{Agostinelli03}. The code simulates the decay of a $^{18}F$ source, the creation of positron, its annihilation with an electron, the generation of the annihilation pair, and the detection of the annihilation photons in the detector ring. A new Geant4 process was created to simulate photon non-colinearity due to positron annihilation-in-flight and non-zero momentum of the electron-positron pair.  

For every photon interaction $i$ within the detector, position $x_i^{G4}$ and energy deposited $E_i^{G4} $  were recorded. 
Interaction sites closer than 1 mm were merged. The detector position resolution was modeled by a Gaussian function (width 1mm). 
 
The number of electrons $N^i_{e-} $ and scintillation photons $S^i $ produced  were calculated as follows
\begin{equation}
\label{eq:charge_el}
N^i_{e-} =  \frac{(1- Fr^*) \times E_i^{G4}}{15.6eV} 
\end{equation} 

 \begin{equation}
\label{eq:light_el}
S^i =  \frac{(SF + Fr^*) \times E_i^{G4}}{15.6eV} 
\end{equation} 
where Fr* is the electron-ion recombination fraction, Fr, with charge-light fluctuation and SF = 0.2 \cite{Aprile07} is the ratio of number of excitons and ion pairs produced.  The charge-light fluctuation was modeled as a Gaussian function centered at Fr = 0.24 with width 0.032 \cite{Astrid09}.
A Gaussian noise (sigma 600 e-) was added to $N^i_{e-} $ to model the electronic noise.  Single interactions producing less than 1800e- were discharged. 
The total number of electrons detected by the LXePET detector was the sum of electrons produced in each single interaction:
\begin{equation}
\label{eq:en_ele}
N^*_{e-} =   \sum_{i=1}^N N^{i*}_{e-}  
\end{equation} 
where N is the number of interactions. 
The APD noise expressed in photoelectrons was modeled as sum of electronic and statistical gain fluctuation contributions as described by \cite{Krishnamoorthy06}: 
\begin{equation}
\label{eq:electr_bl}
\sigma APD_k= \sqrt{(\left(\frac{ENC}{G}\right)^2 + (F-1) \times  N^{APDk}_{pe}}
\end{equation} 
where the number of photoelectrons seen by each APD is given by: 
\begin{equation}
N^{APDk}_{pe} = P(QE \times S_{APDk})
\end{equation} 
and ENC  is the equivalent noise charge (ENC= 1000e-), G is the gain (G=500), F  is the excess noise factor (F=2.5), P(x) is the Poisson function, QE  is the quantum efficiency (QE=$80\%$), and $S_{APDk}$ is the number of scintillation photons reaching each APD.  
In our case the value of $S_{APDk}$ is given by 
\begin{equation}
S_{APDk} = \sum_{i=1}^N S^i \times \phi_{APDk} (x_i^{G4})
\end{equation} 
where $\phi_{APDk}(x)$ is the fraction of solid angle seen by the $APD_k$.

The total number of scintillation photons detected by the LXePET detector corrected for solid angle using the position from charge was calculated as 
\begin{equation}
\label{eq:ScintSpectrum}
S^*= \frac{\sum_{k=1}^{K} N^{APDk*}_{pe} }{\phi(x^G) \times QE}
\end{equation} 
where $ \phi$ is the fraction of solid angle seen by the APDs array, $x^G$ is  the center of gravity of the interaction, and K is the number of APDs. 
Number of electrons (\ref{eq:en_ele}) and scintillation photons (\ref{eq:ScintSpectrum}) detected by the LXePET detector were used to calculate the combined energy as described in \cite{Aprile07}. 
Events with combined energy 450-600 keV were selected and, in case of multiple interactions, their interaction sequence was reconstructed using the Compton reconstruction method \cite{Oberlack00}.  Scattering angles at each interaction point were calculated from energy and geometry. The interaction sequence with the lowest chi-square was assumed to be the correct time sequence.

To evaluate the energy resolution of the PET system we simulated a $^{18}F$ point source embedded in an acrylic cube (side = 1cm) at the center of the
field of view (FOV), whereas for the evaluation of the sensitivity we simulated the source in a polyethylene cylindrical phantom (D=mm, H=4mm) in three different positions along the radial direction (0, 10 mm, and 20 mm) and five different positions along the axial direction (0, 15 mm, 30 mm, 45 mm, and 55 mm).  The sensitivity was calculated as ratio of annihilation pairs with energy 450-600 keV  to emitted annihilation pairs. The active cathode and anode area of the TPC were 32x100 mm and 92x100 mm respectively. The drift length was 11.2 cm. The distance between the two 24 APDs arrays was 110 mm. The axial length was 10 cm. The detector ring diameter was 10 cm. 

\begin{figure}[t]
\centering
\includegraphics[width=16pc]{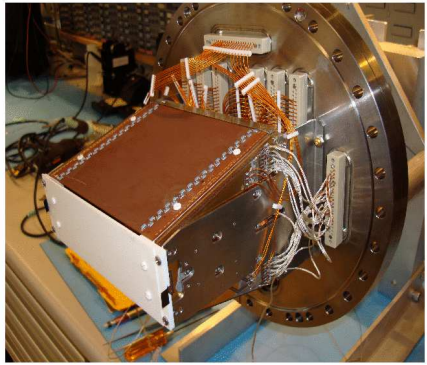}
\caption{\label{fig:photoSector.eps}LXe micro-PET sector mounted on a flange.}
\end{figure}

\section{PET Sector Prototype} 
Cosmic ray muon events were acquired with the LXePET sector to evaluate the detector position resolution. 
The LXePET sector was placed inside a 8.5l cryostat. The active volume of the LXePET sector was 1.1l. Scintillation light was measured by 32 16mm dia. APDs (Advanced Photonix Inc) located on the two sides of the sector.  The anode module consisted of 96 induction wires with 1.1 mm spacing and 96 orthogonal anode strips with 1.1 mm spacing located 1.6 mm above the layer of wires. An electric field of 0.9kV/cm was applied between the cathode and the Frisch grid separated by 12 cm. A field cage provided a uniform electric field in the drift region. The Frisch grid consisted of 30 $\mu$m dia. wires, (cell size= 0.5 mm) located 3.2 mm from the the anode strips. The electric field between the wires and the anode was set to 4 kV/cm to ensure good grid transparency. The detector was operating at 17 psia and at a temperature of 169 K. Figure \ref{fig:photoSector.eps} shows the sector mounted on a flange before its insertion in the cryostat. 

In order to achieve a satisfactory purity level, the vessel containing the LXePET sector was bake-out in vacuum for 7 days at 100\textsuperscript{$\circ$}C and then flushed with hot xenon (70\textsuperscript{$\circ$}C) for 7 days. The xenon purification was performed both in gas (5cc/min) and liquid phase (0.35 L/min).

The signals from strips and wires were sent first to charge-sensitive shaping amplifiers, then to waveform digitizer VF48 modules \cite{Martin06}. The sampling rate was set to  15MHz. The observed noise of the charge-sensitive amplifier was 700 electrons. The APD signals were sent to a current-sensitive preamplifier and to a VF48 digitizer module. The sampling rate was set to 60MHz. The APDs gain was set to 600. The effective electronic noise was 10\textsuperscript{4} electrons.

Cosmic ray muon events triggering at least one APD were recorded. Noisy events, events with multiple tracks, and
events with less than 16 strips and 6 wires were rejected. The accepted tracks were fitted with a linear function in the xz plane. Tracks with $\chi_2 >50 $ or shorter than 8 $\mu s$ were rejected. As result of this selection procedure, 236 muon events were considered. The distance of each hit from the linear fit of the selected tracks was calculated. 

\section{Results and Discussion} 
\subsection{Simulated Performance of the LXe Micro-PET System}
\begin{figure}[t]
\begin{minipage}[b]{16pc}
\includegraphics[width=16pc]{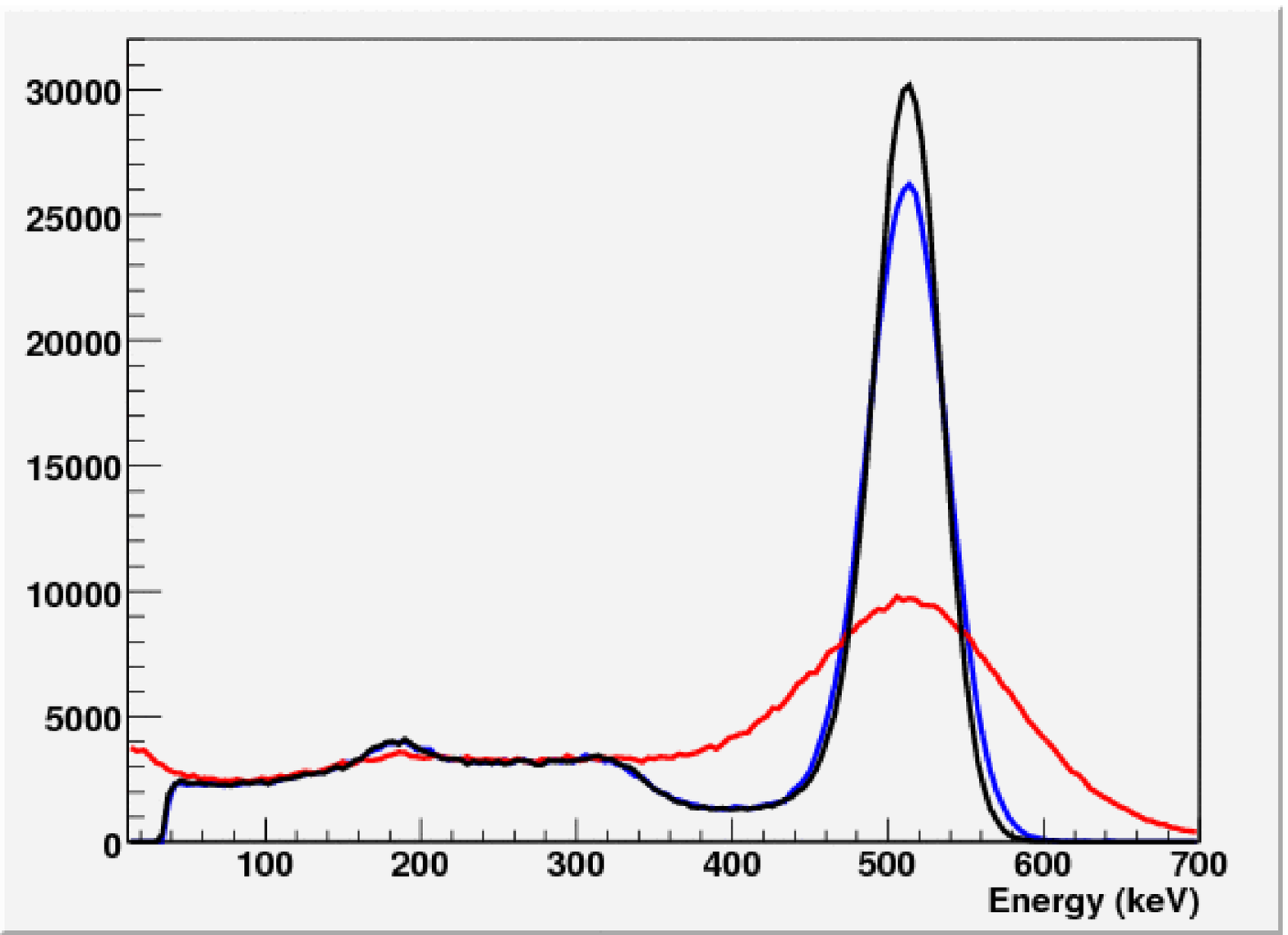}
\caption{\label{fig:spectraSim}Simulated energy spectrum from charge (blue line), light (red line) and combined (black line).}
\end{minipage}  \hspace{1.5pc}
\begin{minipage}[b]{16pc}
\includegraphics[width=16pc]{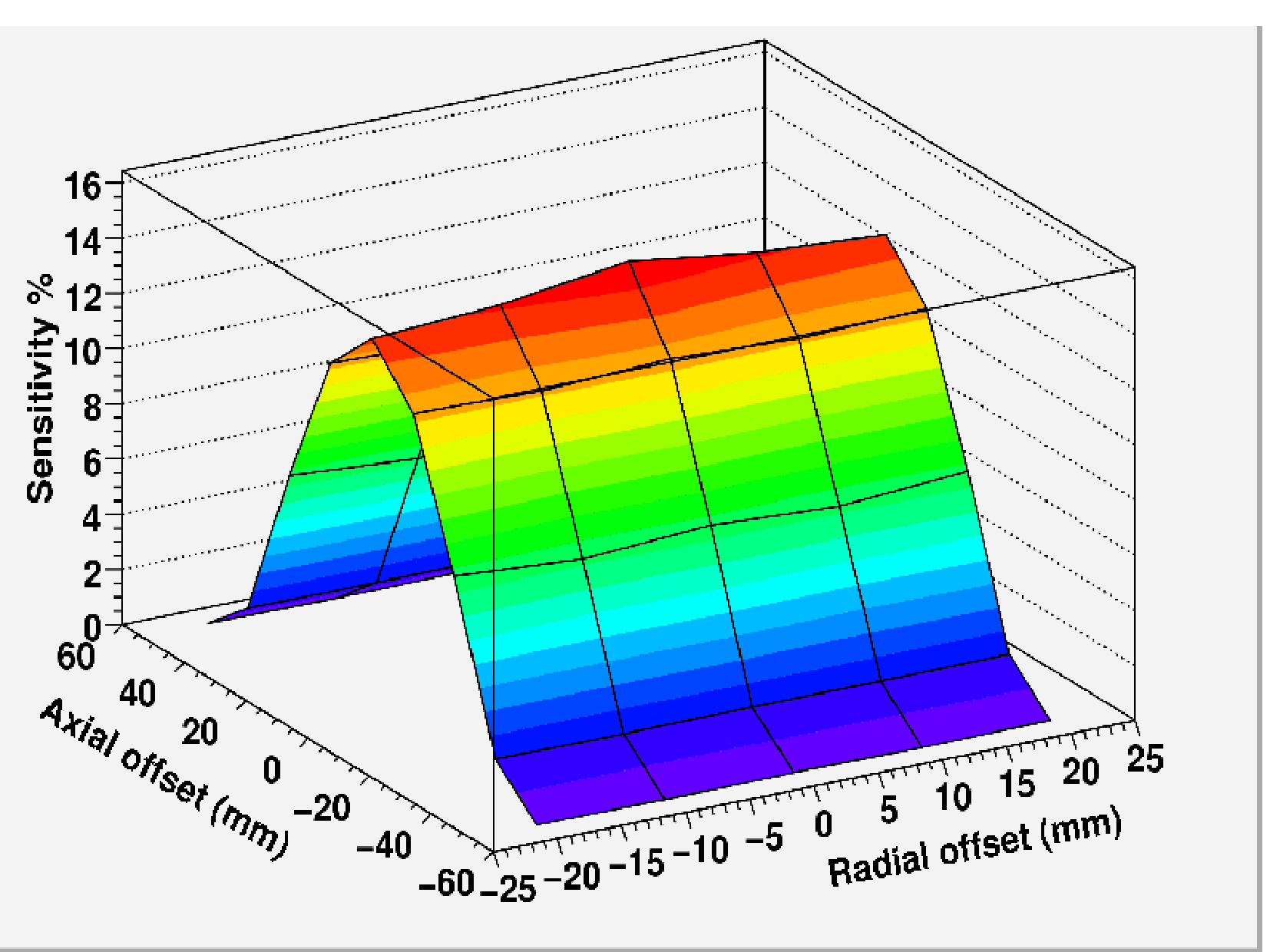}
\caption{\label{fig:simulatedsensitivity}3D sensitivity map of the PET system simulated using a $^{18}F$ radioactive source.}
\end{minipage} 
\end{figure}

Figure \ref{fig:spectraSim} shows the simulated charge, light, and combined energy spectra of a $^{18}F$ point source embedded in a 1cm acrylic cube at 2.66 kV/cm drift field. The resolution from ionization charge, scintillation light, and combined were 11.9$\%$, 32.5$\%$, and 9.9$\%$ (FWHM) respectively. The simulated sensitivity map is shown in Figure \ref{fig:simulatedsensitivity}. The sensitivity at the center of the FOV is 15$\%$ and remains constant along the radial direction, whereas it sharply decreases along the axial direction due to the solid angle.

\begin{figure}[t]
\centering
\includegraphics[width=23pc]{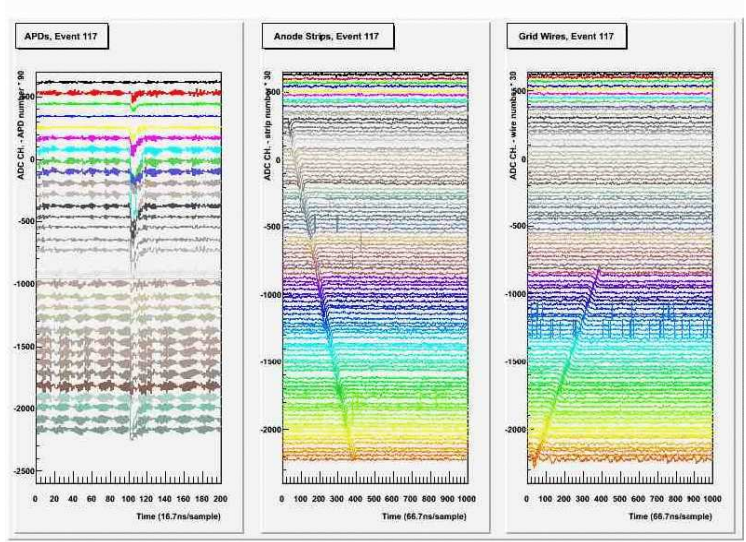}
\caption{\label{fig:muonTrack} Event display showing a cosmic ray muon event.}
\end{figure}

\begin{figure}[t]
\centering
\includegraphics[width=16pc]{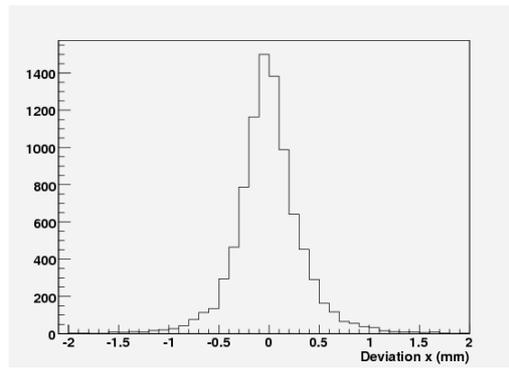}
\caption{\label{fig:residual}Residual in x coordinate for cosmic ray muon tracks.}
\end{figure}

\subsection{First Results with the Test Sector}
A typical cosmic ray muon event is shown in Figure \ref{fig:muonTrack}. 
The digitized waveforms from the APDs (left), anode strips (center),
and induction wires (right) are plotted as a function of time. The waveforms of anode strips and induction wires extending over the full drift time (50 $\mu s$) are sampled every 66.7ns. Deviations of cosmic ray muon tracks from linearity are shown for the x
coordinate in Figure \ref{fig:residual}. The FWHM of the residuals is 0.48 mm.

\section{Conclusions and Future work} 
Preliminary results of the LXe PET sector prototype and simulation results have been encouraging for the application of LXe detectors for PET.
Simulations of the LXePET system including positron range and photon non-colinearity show an energy resolution of $<10 \%$ (FWHM)  and sensitivity
15$\%$ at the center of the FOV.  Tests with cosmic rays indicate a position resolution better than 1 mm. Coincidence tests with a $Na^{22}$ point source and
a BrilLanCe detector are planned for a complete study of the LXePET detector performance. In addition, PET measurements
with two LXe PET detectors will be performed to evaluate image quality, spatial resolution, and sensitivity.

\section*{Acknowledgments}
This work was supported in part by NSERC, CIHR (CHRP Program), the Canada Foundation for Innovation, the University of British Columbia, and TRIUMF which receives federal founding via a contribution agreement through the National Research Council of Canada.

\section*{References}
  
\bibliography{Miceli_INPC2010_ArXiv.bib}

\end{document}